# Ice-templated porous alumina structures

Sylvain Deville[1], Eduardo Saiz[*], Antoni P. Tomsia

*Materials Sciences Division*

*Lawrence Berkeley National Laboratory, Berkeley, CA 94720, USA*

**Abstract**

The formation of regular patterns is a common feature of many solidification processes involving cast materials. We describe here how regular patterns can be obtained in porous alumina by controlling the freezing of ceramic slurries followed by subsequent ice sublimation and sintering, leading to multilayered porous alumina structures with homogeneous and well-defined architecture. We discuss the relationships between the experimental results, the physics of ice and the interaction between inert particles and the solidification front during directional freezing. The anisotropic interface kinetics of ice leads to numerous specific morphologies features in the structure. The structures obtained here could have numerous applications including ceramic filters, biomaterials, and could be the basis for dense multilayered composites after infiltration with a selected second phase.

**Keywords**: porous ceramics, multilayer, freezing, directional solidification, physics of ice.

1. **Introduction**

The formation of regular patterns is a common feature of many solidification processes, such as eutectic growth or unidirectional solidification of two-phases systems [1, 2]. Control of the regularity and size of the patterns is often a key issue with regards to the final properties of the materials. Hence, particular attention has been paid to the control of solidification microstructures in the presence of inert particles [3], both theoretically and experimentally, for its wide application in cast materials. The final microstructure is directly related to the shape

---

and behavior of the solidification front, which can either engulf or repel the inert particles, such as ceramic particles in a solidifying metal.

Porous materials have attracted considerable attention as a new class of materials with a wide range of applications, from bone substitutes to parts for the automotive industry. In these materials, control of the size and morphology of the porosity is often the most critical factor. Cellular ceramics can be engineered to combine several advantages inherent from their architecture [4]: they are lightweight, can have open or closed porosity making them useful as insulators or filters, can withstand high temperatures and exhibit high specific strength, in particular in compression [5]. Typical processing methods include foam or wood replication [6-9], or direct foaming [10].

Between the multiple techniques used to prepare porous ceramics, freeze casting has not attracted too much attention so far, although its simplicity certainly makes it appealing. The technique consists of freezing a liquid suspension (aqueous or not), followed by sublimation of the frozen phase and subsequent sintering, leading to a porous structure with unidirectional channels in the case of unidirectional freezing, where pores are a replica of the ice crystals [2] (in case of aqueous slurries). Although applied to a wide variety of materials so far, like alumina [11, 12], hydroxyapatite [13], silicon nitride [14], NiO-YSZ [15], or polymeric materials [16], a proper and rational control of the microstructure morphology has not yet been achieved.

In freeze-casting, the particles in suspension in the slurry are rejected from the moving solidification front and piled up between the growing columnar or lamellar ice, in a similar way (Fig. 1) to salt and biological organisms entrapped in brine channels in sea ice [17]. The variety of materials processed by freeze casting suggests that the underlying principles of the technique are not strongly dependent on the materials but rely more on physical rather than chemical interactions. The phenomenon is very similar to that of unidirectional solidification of cast materials and binary alloys, with ice playing the role of a fugitive second phase.

The porosity of the sintered materials is a replica of the original ice structure. Since the solidification is often directional, the porous channels run from the bottom to the top of the samples. In addition, the pores exhibit a very anisotropic morphology in the solidification plane. The final porosity content can be tuned by varying the particles content within the slurry, and the size of porosity is affected by the freezing kinetics [11]. More elaborated experimental setups have been designed to obtain radially oriented porosity [15]. The surface of the channels is covered by dendritic-like features, probably related to the morphology of the ice front, though no direct interpretation has been probed so far.

The motivation for this work was therefore to investigate freeze-casting of ceramic slurries, and in particular the relationships between the freezing conditions and the final microstructures, for moderate to highly concentrated suspensions, and interpret the phenomenon in terms of interaction between the solidification front and the inert ceramic particles. We have discovered how under proper control of the freezing conditions, porous multilayered ceramics can be obtained [2]. The experimental setup was inspired by the ones used for the two-dimensional freezing experiments of low concentration solutions [18, 19], and modified for the processing of large three-dimensional samples, which may be characterized from a microstructural point of view. Alumina was used as a model material to investigate the relationships between the freezing conditions and the structure morphology. This inert oxide can be used to prepare stable and well-dispersed aqueous slurries with a wide range of solid contents.

## 2. Experimental techniques

Slurries were prepared by mixing distilled water with a small amount (1 wt% of the powder) of ammonium polymethacrylate anionic dispersant (Darvan C, R. T. Vanderbilt Co., Norwalk, CT), an organic binder (polyvinyl alcohol, 2 wt.% of the powder) and the alumina powder (Ceralox SPA05, Ceralox Div., Condea Vista Co., Tucson, USA) in various proportions. Slurries were ball-milled for 20 hrs with alumina balls and de-aired by stirring in a vacuum desiccator, until complete removal of air bubbles (typically 30 min). The powder used in the study (Fig. 2a) has a specific area of 8.1 $m^2/g$ and an average grain size of 400 nm (data provided by manufacturer). A second powder with smaller particle size (<100 nm) (Fig. 2b) was also used, details can be found in the reference [20].

Freezing of the slurries was done by pouring them into a teflon mold (18 mm diameter, 30 mm length) with two copper rods on each side which were cooled using liquid nitrogen [2] (Fig. 3). Freezing kinetics were controlled by heaters placed on the metallic rods and thermocouples placed on each side of the mold. Freezing occurred from bottom to top of the sample. Frozen samples were freeze-dried (Freeze dryer 8, Labconco, Kansas City, MI) for 24 hours. By adjusting both the temperature gradient and the cooling rate, a wide range of freezing conditions can be investigated. For low freezing rates, only the bottom cold finger was used. To reach higher freezing rates, a constant macroscopic temperature gradient was established using the two cold fingers cooled at the same rate. The average ice front velocity was estimated by measuring the time of freezing and dividing by the length of the frozen sample.

The green bodies thus produced were sintered in air for 2 hours at 1500°C, with heating and cooling rates of 5°C/min (1216BL, CM Furnaces Inc., Bloomfield, NJ). The microstructure of the samples was analyzed by optical and environmental scanning electron microscopy (ESEM, S-4300SE/N, Hitachi, Pleasanton, CA) and their total porosity was derived from the apparent density, measured by Archimedes's method. The wavelength was measured in the top 2/3 of the sample where a constant thickness of the ceramic lamellae was reached. The wavelength was measured on lines perpendicular to the ceramic lamellae. More than 100 measurements per sample were performed.

## 3. Results

### 3.1 General features of the microstructure

If the slurry is partially quenched, i.e., poured over a cold finger maintained at a constant and negative temperature, the initial freezing is not steady. Although lamellae and channels are observed all over the sample, their orientation over the cross-section parallel to the ice front is completely random (Fig. 4a). Colonies of locally aligned pores are observed, but no long-range order is found. Homogeneous freezing (i.e., cooling of the fingers at constant rate starting from room temperature) results in a more homogeneous ice nucleation [21] leading to a lamellar porous architecture (Fig. 4b), with long range order, both in the parallel (Fig. 5a) and perpendicular (Fig. 5b) directions of the ice front. Channels run continuously from the bottom to the top of the sample, originating from a steady state solidification front and previously continuous ice crystals. After sintering, the lamellae (Fig. 6) separating the lamellar channels are completely dense (Fig. 6d) with almost no visible residual porosity. The surface of the lamellae exhibits a particular topography, with dendritic-like features, 1-5 microns high, running in the solidification direction (Fig. 6a-b). These features are homogeneous in size and distribution, but their relative size varies with the freezing conditions. Interestingly, after homogeneous freezing the dendritic surface relief covers only one side of the lamellae (compare Figs. 6a and 6c) while after quenching, dendrites are found on both sides (Fig. 4a). This particular feature is discussed in paragraph 4.3.

**3.2. Influence of cooling rate on the microstructure**

When the freezing kinetics is increased, i.e., the solidification front speed increases, the width of the channels and of the lamellae is drastically affected. As shown in figure 7, the structure wavelength (defined in Fig. 1) can be varied over a wide range, from 7 to 130 µm, for a sample with 64% total porosity; the lamellae thickness in this case varies correspondingly from 2 to 44 µm (Fig. 7). The faster the freezing rate, the finer is the microstructure. The empirical dependence (Fig. 7) between the wavelength ($\lambda$) and the speed of the ice front in the direction parallel to the temperature gradient ($v$) can be described with a simple power law ($\lambda \sim v^{-n}$). The dependence of n with the particle size brings some interesting observations. For the alumina with a 400 nm particle size, $n \approx 1$, while when the particle size decreases to ~100 nm, $n \approx 2/3$ as reported for a simple system with no particles (PVA/water [16]). The behavior of slurries with ceramic nanoparticles tends to get close to that of the simple system solute/water—a system much simpler to model. In addition, preliminary results [2] indicate that this variation might be dependent on the particle content, as pointed out by the previous studies [3, 22].

**3.3 Influence of particles concentration in the slurry**

The microstructure can also be modified by varying the concentration of the starting slurry (Fig. 8). Since the water initially present in the slurry is converted first into ice that is later eliminated to form the porosity, the pore content can be adjusted by tuning the slurry characteristics. The final porosity of the material is directly related to the volume of water in the suspension. Some limits are encountered. At low ceramic content, the green body becomes weaker and difficult to handle. Obtaining samples for ceramic contents lower than 40 wt% is therefore difficult without increasing the binder content, but the features of the lamellar structure do not seem to be affected by the decrease of particle content. On the other hand, when the ceramic content is too high (>80 wt%) (Fig. 8c), the lamellar structure is lost and the pores are not interconnected.

## 4. Discussion

### 4.1 Pattern formation mechanisms: the physics of ice and the interaction with inert particles

In order to obtain ceramic samples with a lamellar porous structure, two requirements must be satisfied:

> (1) The ceramic particles in suspension in the slurry must be rejected from the advancing solidification front and entrapped between the growing ice crystals.
>
> (2) The ice front must have a columnar or lamellar morphology.

Requirement (1) can be easily satisfied, and can be understood in terms of the interaction between an advancing solidification front and inert particles. Various analyses of this phenomenon in model suspensions [3, 19, 23] or biological or natural systems like blood [24] or saltwater [25] have been published. All these fundamental investigations were made for suspensions with a very low concentration of particles, so that the interactions of particles with each other were not taken into account. The thermodynamic condition for a particle to be rejected by the solidification front is that there is an overall increase of free energy if the particle is engulfed by the solid:

$$\Delta\sigma = \sigma_{sp} - (\sigma_{lp} + \sigma_{sl}) > 0 \quad \text{(Eq. 1)}$$

where $\sigma_{sp}$, $\sigma_{lp}$ and $\sigma_{sl}$ are the interfacial free energies associated with the solid-particle, liquid-particle and solid-liquid interface respectively. For slow growth rates and in the absence of external forces, this thermodynamic criterion is often enough to predict if the particle is going to be entrapped or rejected by the solid [19, 26]. However, for the solidification front to push the particles, a liquid film of enough thickness should exist between the solidification front and the particle in order to maintain the transport of molecules towards the growing crystal. When the velocity of the front increases, the thickness of the film decreases. There is a critical velocity, $v_c$, for which this thickness is not enough to allow the necessary flow of molecules to keep the crystal growing behind the particle, that becomes then encapsulated by the solid. Several expressions have been derived for $v_c$, the critical velocity above which a particle or radius *R* will be entrapped. Most expressions define this critical velocity v$_c$ as a function of the particle size R: $v_c \propto \frac{1}{R}$ [24]. A simple model [16, 19] based on the force balance at the particle/ice interface yields a critical ice front velocity for particles entrapment described that can be written as:

$$v_c = \frac{\Delta\sigma d}{3\eta R}\left(\frac{a_0}{d}\right)^z \quad \text{(Eq. 2)}$$

where *a$_0$* is the average intermolecular distance in the film, *d* is overall thickness, η is the solution viscosity, *R* the particle radius and *z* is an exponent that can vary from 1 to 5 depending on the specific model. The main problem with Eq. 2 is to estimate correctly *d, z* and Δ*σ*. The behavior of nanoparticles as the ones used here have never been investigated experimentally, so that the actual critical velocities can only be extrapolated from results obtained with larger particle size. Typical critical velocities vary from 1 to 10 µm/s for particles with a diameter ranging from 1 to 10 µm (e.g., see Fig. 9 in [26]). Since the critical velocity is inversely proportional to the particle radius (Eq. 2), it is likely to be very high (>100 µm/s) for submicronic particles. A very rough estimate of the critical speed can also be obtained by disregarding the correction to the disjoining force of the particle (*z* = 1) and using some reasonable values for the other parameters (R~$10^{-7}$ m, Δ*σ* ~ *σ$_{sl}$* ~ $10^{-2}$ J/m$^2$, *a$_0$* ~ $10^{-8}$ m, η ~ $10^{-2}$-$10^{-3}$ Pa·s) then *v$_c$* ~ 1 - 0.1 m/s. These comparisons support our observation that after the initial fast cooling (see paragraph 4.2) the alumina particles are always rejected by the ice front.

As far as requirement (2) is concerned, two possibilities are to be considered: the planar to columnar transition of the ice front can be generated by the constitutional supercooling ahead of the freezing front, or can be induced by the presence of particles. The crystal structure of ice is such that it has a very low solubility limit for impurities. Therefore, once an ice crystal is formed, any solute initially present in the water will be separated from this growing pure ice crystal and a local increase of solute concentration arises ahead of the ice front. If there is no equilibrium between the diffusion rate of the rejected solute and the rate of crystal growth, a concentration gradient builds up in the zone surrounding the ice front. The concentrated solute lowers locally the freezing point of the solution leading to the formation of a constitutional supercooling zone in an unstable situation, which may eventually lead to the breakdown of the planar interface and the formation of a columnar interface, a phenomenon better known as a Mullins-Serkerka's instability [27]. Further decrease in the ice front velocity (e.g., externally imposed) will results in a columnar to lamellar/dendritic transition. In our aqueous alumina slurries the chemical dispersant and polymeric binder added due to processing requirements are solutes rejected from the growing ice front that can provide the required constitutional supercooling and trigger the transition to the columnar or lamellar/dendritic ice front.

However, the breakdown of the planar interface may also be triggered by the presence of the particles in the liquid [23]. This breakdown can occur for velocities below the threshold for the onset of the Mullins-Sekerka instability [23]. Observation of the microstructure (Fig. 9) suggests that the initial planar ice front is trapping the particles resulting in a dense layer at the bottom of the sample. Afterwards, the formation of a cellular porous microstructure forms indicating that the columnar ice front is rejecting the particles. This will suggest that initial freezing (the initial 10 µm) is very fast and the ice engulfs the particles. When the ice front velocity decreases below $v_c$ the rejected particles pin locally the ice front triggering the transition from planar to columnar (Fig. 9). The ice front velocity parallel to the crystallographic c axis is $10^2$ to $10^3$ times lower than perpendicular to this axis. After the transition to columnar ice occurred, ice platelets with a very large anisotropy can then be formed very fast with ice growing along the *a*-axes, while the thickness (along the *c*-axis) remains small. The freezing process is easier for crystals whose *c*-axes are perpendicular to the temperature gradient, such that growth along the gradient can occur in the *a*- or *b*-direction. The crystals with horizontal *c*-axes will therefore grow at the expense of the others and continue to grow upward, in an architecture composed of long vertical lamellar crystals with horizontal *c*-axes. In the final structures, the direction perpendicular to the lamellae corresponds thus to the original *c*-axis of ice crystals [17]. A better understanding of the pattern formation in this case might be explained by incorporating the effect of the particles and the constitutional supercooling. Segregation of solutes will also affect the interfacial energies and in consequence the critical velocity for particle entrapment. Modeling of this complex phenomenon is well beyond the scope of this paper. However, in our preliminary experiments different solutes do not strongly affect the time of the planar to columnar transition suggesting that the particles play a crucial role [28].

### 4.2 Initial gradient and morphological transition

Independently of the phenomenon triggering the morphological transition, some time is necessary for the freezing front to move from a planar to a columnar or lamellar/dendritic morphology. The transition is reflected in the final architecture of the porous structures, in the first frozen zone. A combination of parallel and perpendicular cross-section reveals the progressive transition and the intermediate stages (Fig. 9). Initial freezing is very fast and then the velocity of the liquid front decreases rapidly until it reaches a steady state with an approximate constant value. Consequently, the first frozen zone reveals a planar ice front where the alumina particles were entrapped. The interface then moves progressively to a columnar and eventually lamellar morphology, with a progressive ordering of the lamellae. A steady state is eventually reached and ice crystals become continuous, running through the entire sample, with a constant thickness. It is worth pointing out here that these observations allow rationalizing previous results of the literature, where morphological transitions during freezing were observed but not explained [29, 30].

### 4.3 Anisotropy of interface kinetics

The thickness of the ice crystals is strongly dependent on the speed of the solidification front. Faster freezing velocities result in larger supercooling in front of the growing crystals that will influence the crystal thickness. In addition, as faster growth is imposed in the direction of the temperature gradient, lateral growth along the c-axis is increasingly limited resulting in thinner lamellae. The results suggest that addition of particles above some critical size of few hundred nanometers will also affect the relative kinetics of crystal growth.

In a two-dimensional cross-section perpendicular to the ice lamellae and the freezing front, it is possible to define two growth directions for the ice crystals: parallel to the temperature gradient (with a growth rate $v$), and along the preferred growth direction (in the sense of interfacial energies), with a growth rate $v_p$. They can be different, this may result in tilted crystal growth (Fig. 10). The balance between imposed and preferred growth results in two characteristic phenomenon: the formation of dendrites on the lamellar ice crystals and tilting of the crystals at very fast cooling rates [31, 32]. The freezing kinetics depends on a myriad of factors (heat of fusion, convection in the suspension, thermal diffusivities…) and is an extremely complex phenomenon to model. However, the ratio $v_p/v$ is strongly dependent on the magnitude of the temperature gradient: $(T_1-T_2)/L$, where $T_1(t)$ and $T_2(t)$ are the time-dependent temperatures of the top and the bottom cold finger and L is the length of the mold that influences the growth kinetics. Two main regimes can be observed in our experiments.

During steady state growth, the temperature gradient and the driving force for directional growth are relatively large and consequently $v \gg v_p$. The ice crystals grow just a few degrees tilted with respect to the direction of the temperature gradient while in one side of the crystals small dendrites few microns in height aligned with the direction of the gradient form. Due to the growth pattern these dendrites are found only on one side of the lamellae (Fig. 6a), the other one being flat (Fig. 6c). The dendrites are observed independently of the initial powder content within the slurry. The ceramics particles entrapped in the channels between the converging secondary fronts of the ice cells, such as the surface dendrites, generate the observed roughness on one side of the ceramic lamellae, while the other remains flat after removing the ice [25]. Modifying the topography of the wall surface will therefore involve a control of the dendritic ice morphology.

As the temperature gradient decreases, the dependence to the imposed growth direction becomes less marked and when the temperature gradient direction and the preferred growth direction (due to crystalline anisotropy) are different the effect of anisotropic interface kinetics may arise. A decrease in the temperature gradient is accompanied by a decrease in the driving force for directional growth and consequently in $v/v_p$. The macroscopic result of this mismatch is an increasing tilt of the ice crystals (and hence the resulting ceramic lamellae, Fig. 11). The overall tilting of the cells depends on the magnitude of the anisotropic interface kinetic effect. This behavior is observed in the first frozen zone of samples prepared using only the bottom cold finger. When this zone forms, the macroscopic gradient is still relatively small (< 1 °C/mm, and the first formed lamellae exhibit a strong tilting (Fig. 9, vertical cross-section), similar to that observed in figure 11. As the temperature of the bottom cold finger decreases, the gradient increases, and a steady freezing state is progressively reached and tilting becomes less marked; the ice crystals become closely aligned along the temperature gradient direction.

**4.4 Ceramic bridges**

Another microstructural feature are the trans-lamellar ceramic bridges (e.g., Fig. 12), observed only for highly concentrated slurries. In the sintered porous structures, these numerous fine features with often-tortuous morphologies are locally bridging the gap between two adjacent lamellae. The morphology of these features is very different to that of the dendrites (see Fig. 6b) covering the ceramics lamellae, suggesting another formation mechanism. We propose here that they might be formed because of the specific conditions encountered during the slow freezing of highly concentrated solutions. The interaction of inert particles and a moving solidification front has been investigated for suspensions with low particles content. In such a case, the interaction between particles is not taken into account, which considerably simplify the associated formalism. In the case of highly concentrated solutions, the particle-particle interactions cannot be ignored anymore. Eventually, it might considerably affect the pattern formation mechanisms. It has previously been shown that the particles themselves may induce morphological transitions, such as dendrites tip splitting or healing during growth before being captured [3]. Ceramic bridges between lamellae may arise from local ice crystal tip splitting and engulfment of particle agglomerates created by particles repelled from the ice-water interface and subsequent tip healing. Depending on the magnitude of tip splitting/healing, the entrapped ceramic particles might not bridge completely the gap.

**4.5 Controlling the structure**

Freeze casting is based on a physical process, the formation of ice, and the results described here for alumina can be extrapolated to a wide range of materials, opening the path towards numerous applications. Recent works have shown how this process can be adapted to the fabrication of porous polymer or hydroxyapatite scaffolds for tissue engineering [13, 16] or as the base for the fabrication of hybrid materials [33] or multilayered nacre-like composites after subsequent infiltration with a suitable second phase [2].

From the previous discussion, it is clear that the relationships between ice front velocity and structure wavelength is extremely complex to describe with a simple analytical model. The investigated system is a multicomponent one, combining the effect of the particles/ice front interactions and solute segregation. The structure wavelength (or ice crystals radius) will therefore depend on a large number of factors, including the freezing rate (or ice front velocity), the interfacial free energy between the particles, the water and the ice front, the particles size, distribution and content, the interactions of the particles with themselves, the anisotropic effects of the surface tension of ice, the buoyancy forces acting on the particles, the viscosity of the slurry, diffusion of the solute away from the interface, latent heat diffusion, etc… Nevertheless, some trends, summarized in figure 13, emerge from our studies.

The empirical dependence (Fig. 7) of the structure wavelength with the particle size brings some interesting observations. It seems that when the particle size becomes negligible in regards of the ice tip radius (~100 nm in our experiment), the behavior of the multicomponent system is approximately the one of the simple system of solute in water, which can be predicted by linear stability analysis. Pursuing structure with thin lamellae will involve two conflicting behaviors. On the other hand, the lamellae thickness cannot be made thinner than the particle size, but by using nanoparticles the exponent $n$ that marks the dependence of the wavelength with the vertical velocity of the ice front decreases and thicker lamellae are obtained. A compromise has to be found in terms of particle size for sizing down the lamellae thickness. There seems to be an optimum in particle size where the structure wavelength reaches a minimum (and hence the exponent $n$ a maximum). Our preliminary results using different additives [28] suggest that they do not have a strong effect on the dependence of the wall thickness with the freezing rate. Investigations will additional well-controlled powders will be needed to clarify these points and identify this optimum.

## 5. Conclusions

Based on the experimental investigations of the controlled freezing of moderate to highly concentrated ceramic aqueous suspensions, the following conclusions can be drawn:

1. Homogeneous lamellar porous structures can arise from the controlled freezing of ceramic slurries, followed by sublimation of the ice and sintering of the porous green bodies. The porosity is open, unidirectional and homogeneous throughout the whole sample.

2. The pattern formation mechanisms can be qualitatively understood by the application of simple principles of the physics of ice, and the interaction of inert particles and the solidification front.

3. The morphology of the porous structures, i.e., the content, dimensions, shape and orientation of porosity can be controlled to some extent by the initial slurry compositions and the freezing conditions. Regarding the highly concentrated solutions, the particle-particle interactions should probably be taken into account to explain the experimental results, in particular the formation of ceramic bridges between lamellae.

Figure Captions

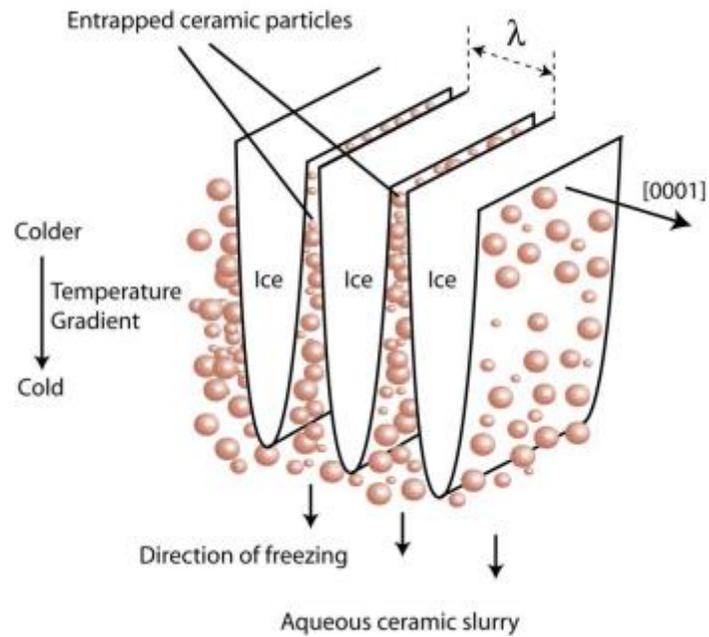

**Figure 1**: Pattern formation and particles segregation during freeze casting of ceramic slurries. The ice platelets grow in a direction perpendicular to the c-axes of hexagonal ice. The wavelength of the structure is defined by λ.

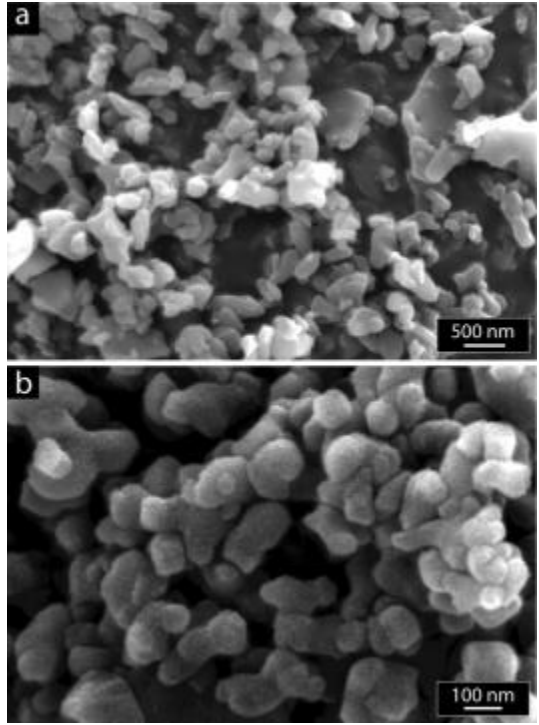

**Figure 2**: SEM micrographs of the ceramic powders used in this study: (a) Ceralox SPA and (b) Fine (~100 nm) alumina powder prepared according to reference [17].

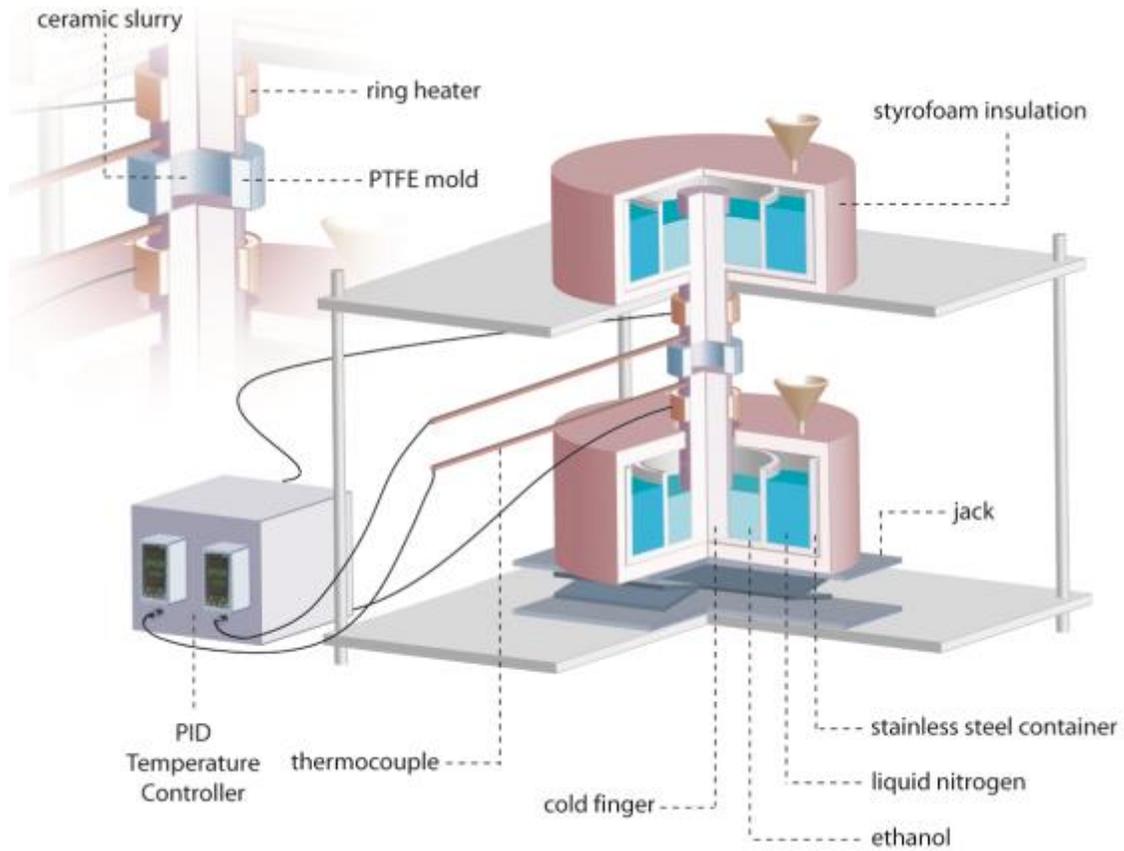

**Figure 3**: Schematic of the experimental apparatus employed to directionally freeze the ceramic slurries while controlling the speed of the freezing front. The ceramic slurry is poured into a Teflon mold placed between two copper cold fingers whose temperature is controlled by liquid nitrogen baths and ring heaters.

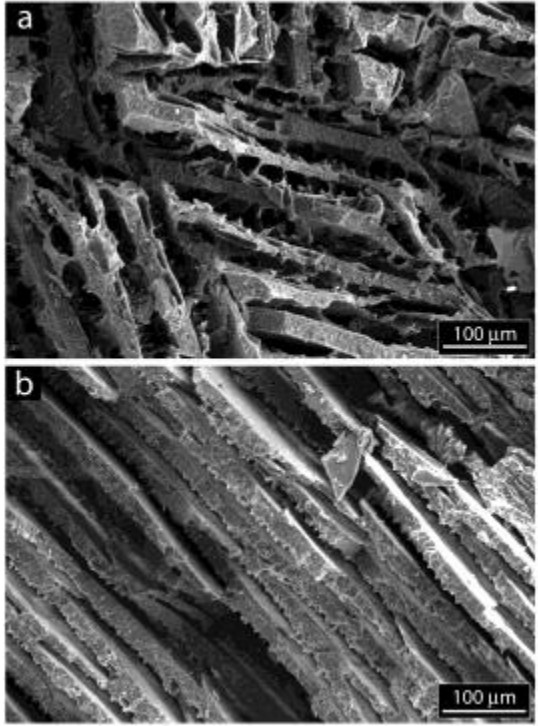

**Figure 4**: SEM micrographs of the porous structure: (a) Isotropic vs. (b) Anisotropic microstructure. Cross-section parallel to ice front.

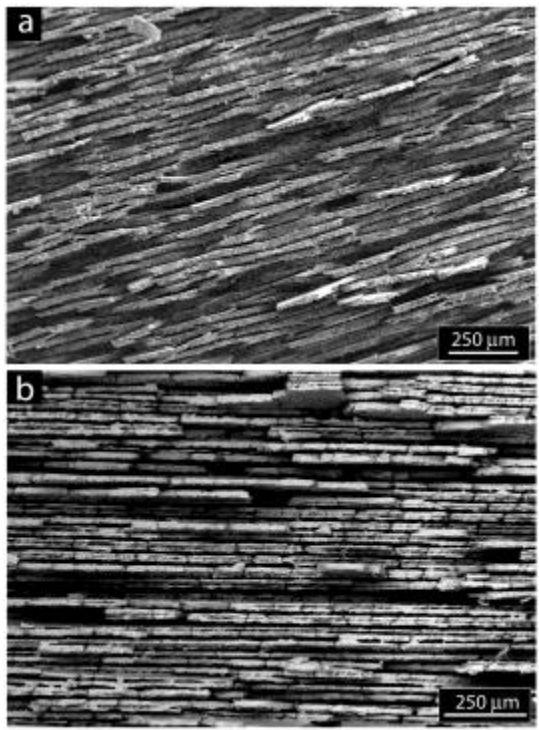

**Figure 5**: SEM micrographs of homogeneous lamellar alumina cross-sections. (a) Parallel to the ice front (b) Perpendicular to the ice front.

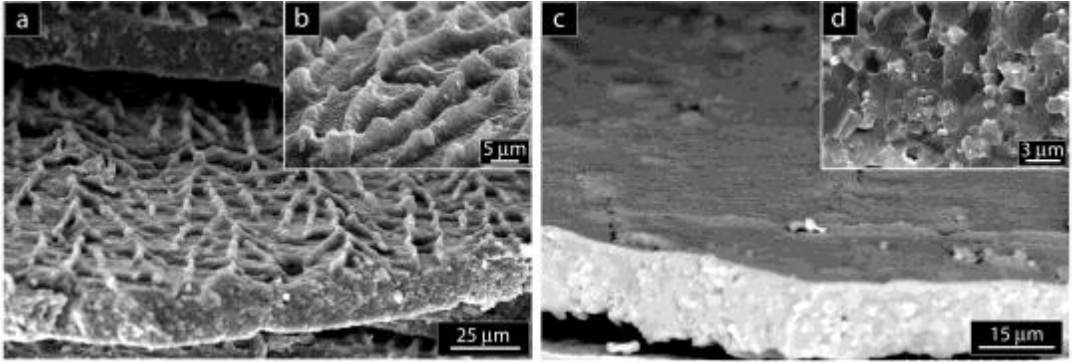

**Figure 6**: SEM micrographs of lamellae surface features: (a) Rough side with (b) Details in insert and (c) Smooth side. (d) Detail of the fracture surface of a lamella, the microstructure shows almost no residual porosity. The microstructure obtained after controlled solidification reveals that only one side of the lamellae is covered with dendritic-like features, the other side remaining flat. This behavior is related to the anisotropic interface kinetics.

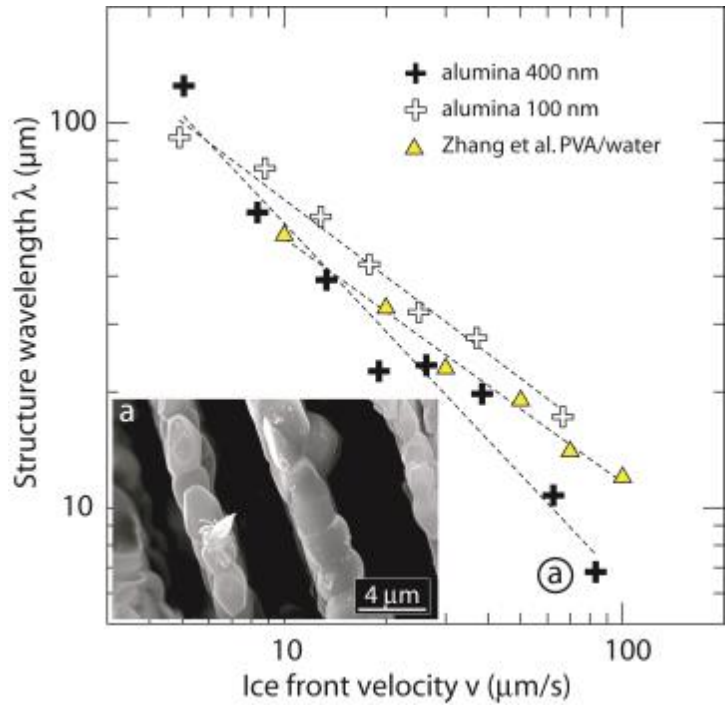

**Figure 7**: Variation of structure wavelength vs. ice front velocity, for samples with 64 % total porosity. The samples were frozen using only the bottom cold finger whose temperature was decreased at a constant rate (ranging between 0.1 and 10°C/min). The only exception is the sample with ~2 µm lamellae thickness where both the top and bottom cold fingers were cooled at 2 °C/min in order to maintain a constant temperature gradient of ~1 °C/cm. The plots can be fitted with power laws, with exponents of 1 (alumina 400nm) and 2/3 (alumina 100nm and PVA/water). Comparison of the plots is discussed in the final section. A SEM micrograph of the sample with the smallest lamellae thickness achieved in this study is shown in the insert (a), cross-section parallel to the ice front. Lamellae thickness is ~ 2μm. Each value on the graph is an average of >100 measurement of wall thickness in cross section, using SEM micrographs. Error bars were not added on the graph since on a log scale, error bars were smaller than the symbols.

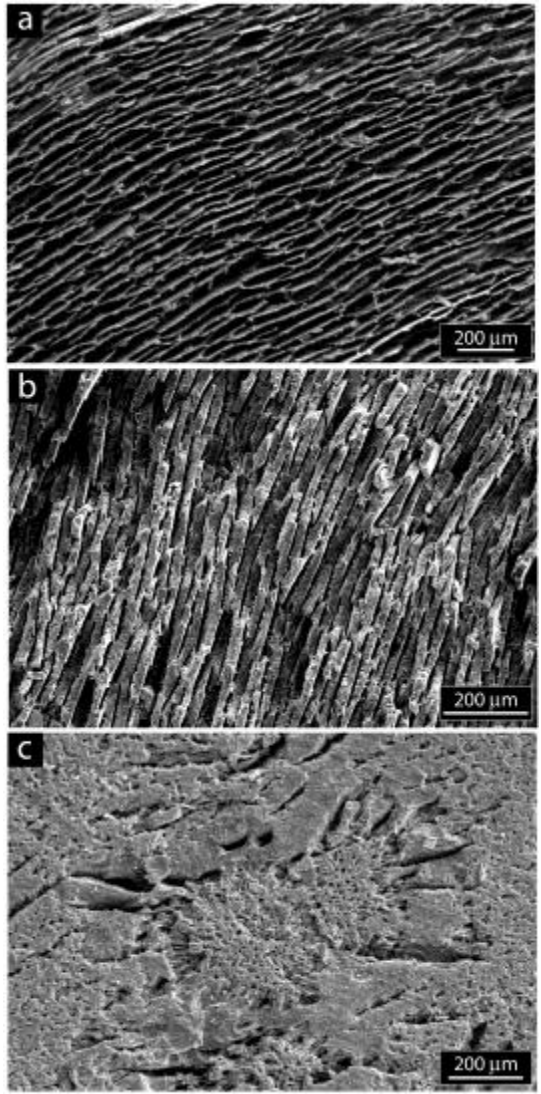

**Figure 8**: SEM micrograph showing the influence of water content in the slurry on the microstructure. Cross-section parallel to ice front. Total porosity of (a) 70 % (b) 40 % (c) 24 %.

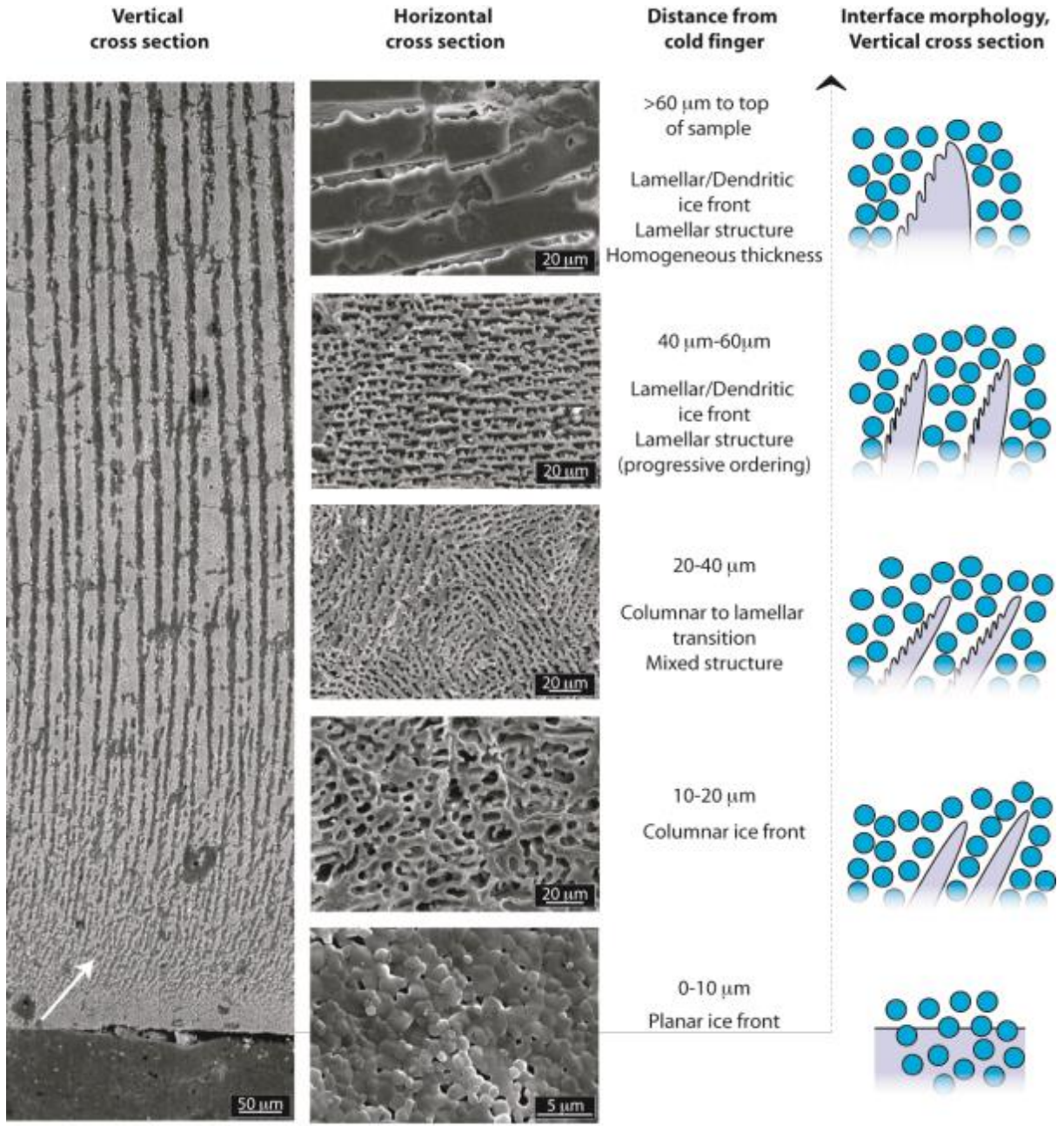

**Figure 9**: SEM micrograph of the final microstructure and evolution of the ice front morphology. The black portion at the bottom of the sample is the epoxy that was used to embed the sample for cross-sectioning and polishing. It takes about 200-250 microns to reach homogeneous layer thickness; the layer thickens progressively with the height and then becomes constant. Tilting of the lamellae in the first frozen zone can be observed at the bottom of the micrograph (white arrow). The horizontal cross-sections (parallel to ice front)

reveal the corresponding evolution of the porous structure and hence interface morphology depicted on the right).

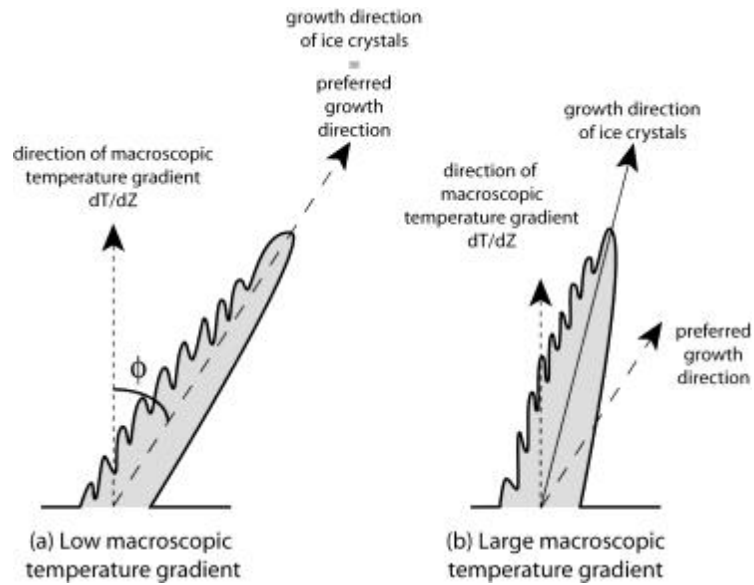

**Figure 10**: Growth direction of ice crystals and relationships with imposed macroscopic temperature gradient and preferred growth direction.

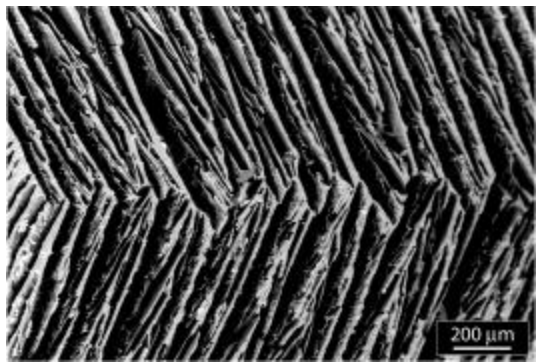

**Figure 11**: SEM micrograph showing tilting of the lamellae due to the effect of anisotropic interface kinetics. Cross-section perpendicular to ice front. In this case, the sample was frozen very fast using the top and bottom cold fingers to set a very small gradient, so that the crystals started on both sides almost simultaneously and joined in the middle of the sample. The tilting (15-20°) is clearly visible.

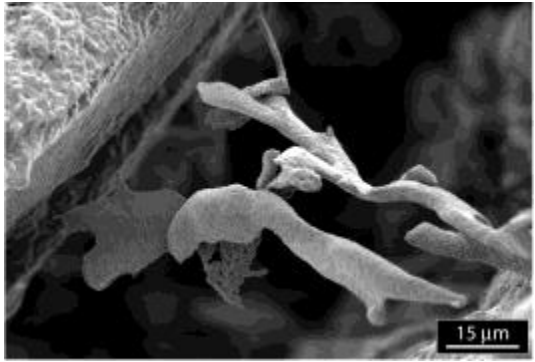

**Figure 12**: SEM micrograph of ceramic bridges observed for slow freezing of highly concentrated slurries.

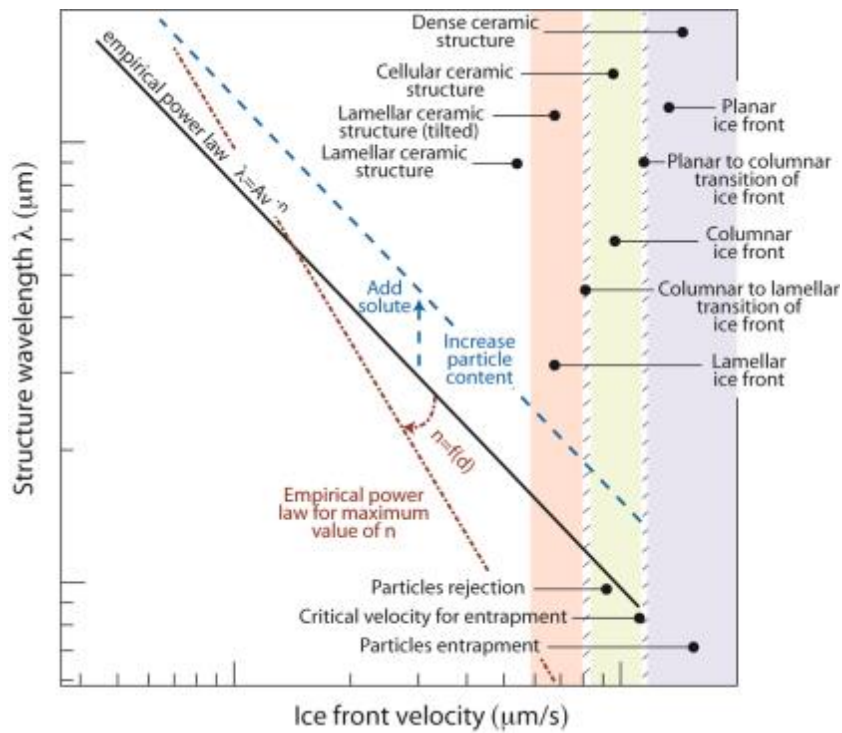

**Figure 13**: Strategies and limits for controlling the structure: schematic plot of wavelength vs. ice front velocity. The exponent n of the empirical law is dependent on the particle size *d*, though the function $n = f(d)$ is not monotonic; an optimum of *d* is encountered where the exponent *n* is maximum. Very fast cooling rates, region (a), will result in the ice front trapping the particles and the formation of a dense material as the one observed in the bottom layer of Figure 9. When the velocity is decreased below the critical value for particle entrapment, $v_c$, the particles are expelled from the growing ice but if the speeds are fast enough the ice will grow with a columnar microstructure, region (b). Slower velocities will result in the formation of lamellar ice. However, if the velocity is still fast or, equivalently, the gradient in temperature small enough, the balance between the preferential growth direction and the gradient direction will result in the growth of lamellae tilted with respect to the later, region (c).  As the velocity decreases (or the gradient increases) the lamellae will align with the direction of the temperature gradient, region (d).